# Toward Cloud Computing Evolution: Efficiency vs Trendy vs Security

Heru Susanto[1,2], Mohammad Nabil Almunawar[1], and Chen Chin Kang[1]

[1]**University of Brunei** - FBEPS, Information System Group. susanto.net@gmail.com & nabil.almunawar@ubd.edu.bn
[2]**The Indonesian Institute of Sciences** - Information Security & IT Governance Research Group. heru.susanto@lipi.go.id

*abstract*-Information Technology (IT) shaped the success of organizations, giving them a solid foundation that increases both their level of efficiency as well as productivity. The computing industry is witnessing a paradigm shift in the way computing is performed worldwide. There is a growing awareness among consumers and enterprises to access their IT resources extensively through a "utility" model known as "cloud computing." Cloud computing was initially rooted in distributed grid-based computing. It has become a significant technology trend and expect that cloud computing will reshape IT processes and the IT marketplace. With the cloud computing technology, clients use a variety of smart mobile devices to access programs, storage, and application-development platforms over the network and internet, through services offered by cloud computing providers.An innovative new way to boost capacity and add capabilities in computing without spending money on a new infrastructure, training new personnel or licensing software is one of the characteristics of a cloud computing. Demands for fastest access to information is changing and increasing, therefore, the availability of cloud computing has made it easier for organizations to share and store related data and information with their stakeholders.Moreover, cloud computing is the use of internet-based services to support business processes.Our research is to find out what the demand and main emphasized of cloud computing compared with efficiency, trendy and security that leads to improved business processes within corporate as cloud user.

*Keyword–*
*Cloud Computing, Cloud Evolution, Cloud Pros-Cons, Cloud Security*

I. INTRODUCTION

Technology continues to advance every day. As technology changes, digital literacy change more rapidly and by keeping up with these variations leads us computer literate. Several years ago, 10 Mbps shared ethernetlocalareanetwork commonly used, followed by switched ethernet and 100Mbps fastethernet. Wireless LAN then introduced with speed up to 54Mbps. The fast prompted of high-speed system developed with new applications such as image and video data. Moreover, the cloud computing which application continues to increase and make work performing easily without software installation on the computer is introduced. Cloud computing has made a long journey since the year 1966 where Parkhill wrote out his ideas in book titled "The Challenge of the Computer Utility" whereby almost all the characteristics of today's cloud computing. Therefore, the platform idea of cloud computing first began during the 1960's by McCarthy, he proposed that in the future "computers may someday be organized as a public utility". He believed with a new way of organizing information and data is only within arm's reach. ServerMotion (2012) mentioned that in the past, massive computing ran within supercomputers and mainframes. They need supercomputers and mainframes to run, manage and organize computing tasks within an organization.

*"...Cloud Computing is a completely new IT technology and it is known as the third revolution after PC and Internet in IT. To be more specific, Cloud Computing is the improvement of Distributed Computing, Parallel Computing, Grid Computing and Distributed Databases,and the basic principle of Cloud Computing is making tasks distributed in large numbers of distributed computers but not in local computers or remote servers. In other words, by collecting large quantities of information and resources stored in personal computers, mobile phones and other equipment, Cloud Computing is capable of integrating them and putting them on the public cloud for serving users...."* (Sanchati and Kulkami, 2011).

Cloud computing software might be more stable and reliable than the software runs on desktop machine (Li, 2009). In the era of modernization, technology has rather become a major need everywhere. Without technology, however businesses, schools, colleges, medical outlets and almost every possible institution will not be able to perform its tasks. Everyone is now relying on technology and there has always been an increase demand for better technology. It was due to the advancement in technology that cloud computing was introduced. Consumers and businesses allowed using applications without installation and they can access personal files on any computer or portable devices such as, Ipad or Iphone, through internet conectivity. Yahoo email, Gmail, or Hotmail are simple examples of cloud computing. However, even though this technology is convenient and cost saving to use, there are several drawbacks and certain restrictions. In 1999, Salesforce.com is one of the first to invest in cloud computing where they introduced the concept of delivering enterprise applications through a simple website,then Amazon launched the Amazon web Service. Then in 2006, Google Docs, which has spread the word of cloud computing and became the leader of public awareness. It followed by collaboration of industries such as between Google, IBM and several universities. With the continuous improvement in computer technologies it will also provide improvement in cloud computing.

Our researchis primarily aimed toidentify the cloud computing evolution and the growth of cloud and the need of technology, within business environment. It is obvious that everyone and businesses is now relying on technology and there has always been an increase demand for better technology, it was due to the advancement in technology that cloud computing was introduced. This paper is organized as follows. In the section 2, we discuss several information technology issues and its contribution as agent of change in business through cloud computing concept. Features, model and current trend issues in cloud computing is discussed in Section 3.Section4, we discussed severalcloud computing phenomenon, such as; benefits from cloud computing: EFFICIENCY, thedrawbacks of cloud computing: SECURITY, the growth of cloud and the need of technology: TRENDY AND DEMAND, and future of cloud computing.Finally, conclusion remarkand future work are provided in Section 5.

## II.    LITERATURE REVIEW

## TECHNOLOGY

Vouk (2008) stated "*today, almost any business or major activity uses or relies in some form, on IT and IT services.*" Cloud computing is one of the IT services that is gaining popularity amongst stakeholderwithin business environment. According to the New York Times (2012), Facebook, a cloud computing software, is the "world's largest social network". Facebook also defined as "a data processor, archiving and analyzingshared of information, from our interests, our locations, to every article and link that we 'like'" (figure 1). However, the main problem of sharing information on the internet is the level of security deal with (Susanto, Almunawar, Tuan, 2012). In 2010, is the timeline of cloud computing, installing web application in the cloud-computing atmosphere becomes the choice of most organizations. Unfortunatelly, Microsoft APAC quotes that only 49% is planning to use cloud in future. Small-medium businesses do not realize that they are currently using this technology. Globally, approximately 370 million people use Hotmail and 310 million use windows livemessenger.

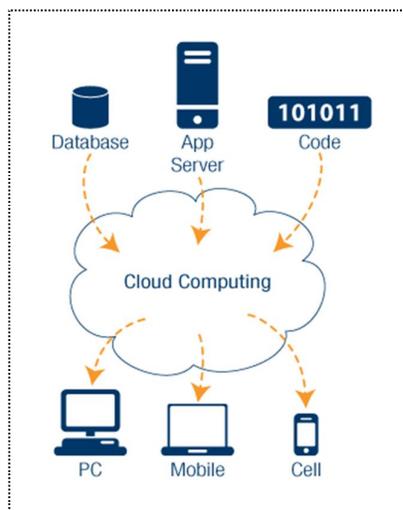

**Figure 1**.*Cloud Computing Scheme and Concept*
*(source: http://www.accessus.net/business-services/cloud-computing/)*

Interactive Cloud Streaming is one of the new technologies developed such as Active Video, GaiKai, GameStop, Onlive and Otoy. The Cloud Streaming gives opportunities to smart portable devices though it experience dispute such as latency and required special video browsers for dissimilar kind of portable devices. The cloud-based real-time complex-event-processing technology created by Fujitsu is another example of cloud computing. It involved process of distributing and paralleling to make it possible for large amount of granularity in handling processed data. This allows on-the-fly alteration to data-load variations.  The lowest effect of processing and jobs that are quiet related are possibly directed to the same server; these are the concern of the new technology when deciding which workload to alter. This technology is developed to make it possible increases in throughput, immediate reply and quicker alteration to changing data loads. Windows Azure is a latest version of Windows where Microsoft is focusing on developing the cloud as a platform or software as a platform.

## CASE STUDY : FACEBOOK

Cloud Computing, is the sharing or storage by users of their own information on remote servers operated by others and accessed through the internet or other connections such as data storage sites, video sites, tax preparation sites, personal health record websites, photography websites, and social networking sites (Cloud Computing and Privacy, n.d.). It is therefore not surprising that Facebook, one the earliest and successful online social network (Ellison, Steinfield&Lempe, 2007) in the world, is also an earlier form of cloud computing. It has a large impact on a wide number of issues (Yadav, 2006).

How Facebook is considered as cloud computing though, is quite easy and simple to understand. Here are some of the main characteristics Facebook has to be considered as a 'Cloud Computing' service: The users may 'own' their Facebook page, which means that they are the ones responsible for the content of their page, whereby, the actual site, original software and hardware are fully and utterly owned by Facebook.
Users added contents to their own 'Facebook' page, which usually includes information, ideas, planned activities and thoughts on severaltopic. Also, Facebook allows its users to upload photos and share videos with others. However, for educational or business purpose, users can also create group page and add selected members in the group to discuss or share information with selected members. For instance, in Universities most lecturers create a group page and add their students to the page and upload documents, perform discussions and share ideas with the students through this group page. Therefore, all documents are stored on Facebook and accessed by members anytime.

Users could develop their own application, widgets, tools and projects based upon Facebook application called "Facebook Developers" (Yadav, 2006), gives users access to Facebook's own virtual infrastructure and attempt to improve it. The responsibility of the 'Facebook page' is by the owner. It mean that whatever information, events or anything the users decide to share through their own Facebook are not responsible for it nor is he responsible for any potential consequences.

## III. FEATURES AND TYPES OF SERVICE MODELS

"...*Cloud computing is an architectural model for deploying and accessing computer facilities via the Internet. A cloud service provider would supply ubiquitous access through a web browser to software services executed in a cloud data center. The software would satisfy consumer and business needs*" (Katzan, 2010). In simple terms, anything that involves using the internet to deliver hosted servicesis considered as Cloud Computing. It is also the presentation of computing towards a service rather than a product where shared data from software, information and to resources are also provided to computers and other platform of devices as a utility over a network that is mainly the Internet.

The National Institute of Standards and Technology (NIST) statedcloud computingas "*a model for enabling convenient, on-demand network access to a shared pool of configurable computing resources (e.g., networks, servers, storage, applications, and services) that can be rapidly provisioned and released with minimal management effort or service provider interaction. This cloud model promotes availability and is composed of five essential characteristics, three service models, and four deployment models*" (Mell & Grance, 2009). The four service deployment models are:

### 1. PUBLIC CLOUD

The cloud infrastructure is open to the publicand controlledby an organization as cloud service provider, which is available on a commercial basis. It makes customerenables to extend and organize a service in the cloud with low cost compared to the cost required compared to other deployment options. However, it makes free publicclouds leads lowersecured than fee paying cloud models, because fee payingmodelmore focus on security inall applications layer and data accessed.

### 2. PRIVATE CLOUD

The cloud infrastructure deployed, operated and maintained bycloud service providerfor an organization special purpose as a client. The security is more better than public cloud, since private cloud as special treatment and services to the client through private network, which is indicated more secure from unauthorized access, use, disclosure, disruption, modification, perusal, inspection, recording or destructionof information.

3. HYBRID CLOUD

Hybrid cloud consists of any clouds type, as feature asclouds have the ability through their interfaces to allow data and/or applications moved from one cloud to another. "Hybrid Cloud provides more secure control of the data and applications and allows various parties to access information over the internet. It has an open architecture that allows interfaces with other management systems. Hybrid cloud describe combining a local device configuration, such as a plug computer with cloud services, describe configurations combining virtual and physical, collocated assets -for example, a mostly virtualized environment that requires physical servers, routers, or other hardware such as a network appliance acting as a firewall or spam filter." (Kuyoro S. O., Ibikunle F., Awodele O., 2011).

4. COMMUNITY CLOUD

Community cloud consists of number an organization and supports specific community concerns (e.g., mission, security requirements, policy, and compliance considerations),it managed by the third party (TWP, 2010). However, there are three cloud service models give a view of what a cloud service is. A cloud service system is a set of elements that facilitated the development of cloud applications (Youseff, 2009).

The three main service delivery models are:

1. INFRASTRUCTURE-AS-A-SERVICE (IAAS)

The clientorganized the operating systems, applications, storage, and network connectivity, but do not control the cloud infrastructure (TWP, 2010). Katzan (2010) stated thatautorized is provide to the consumer for provision processing, storage, networks, and other fundamental computing resources where the consumer is able to set up and run arbitrary software, included operating systems and applications. Thus, the clientdoes not organize underlying cloud infrastructure but control over operating systems, storage, deployed applications, and possibly limited control of select networking components (e.g., host firewalls).

2. PLATFORM-AS-A-SERVICE (PAAS)

The clienthas authorityaccess to the platforms, enabling to deploy their own software and applications in the cloud (TWP, 2009). Unfortunately,operating systems and network access are not managed by the consumer, and there might be constraints as to which applications can be deployed,itmeansclients is ableto deploy onto the cloud infrastructure consumer-created or acquired applications created using programming languages and tools supported by the provider (Katzan,2010)

3. SOFTWARE-AS-A-SERVICE (SAAS)

Consumers are purchase the ability to access and use an application or service hosted in the cloud (TWP, 2010),as an example is Salesforce.com, where crucial information for the relations between the consumer and the service hosted as part of the service in the cloud. Katzan (2010), the capability provided to the consumer is to deploy onto the cloud infrastructure consumer-created or acquired applications created using programming languages and tools supported by the provider. The clientdoes not organize the underlying cloud infrastructure, but has control over the deployed applications and possibly application hosting environment configurations.

## IV. Cloud Computing Phenomenon

### Benefits from Cloud Computing: Efficiency

The Gates Foundation reveals that invested in training for tutors is a way to develop learning for school learner (Brad, 2011). Cloud computing provides convenience to the users due to its ability to access the applications anywhere as long as there is an internet connection (Diane, 2011). In business world, cloud computing is suitable for storage facilities as it has unlimited storage capacity. The users have no worry if their computer crashes or is out of service since data is still accessible in the cloud. Therefore, the users can access data from different portable devices. Scalability refers to the change in measurement of the data storage. As the business grows, it is possible to get additional license and storage space. It is easier to manage as users have no worry about maintaining or updating the software and hardware. In addition, most of the clouds computing application software are free and it is much better in terms of performance, as compared to the expensive application software.

Cloud computing assissted stakeholderand businesses procesess to saves budget and offered efficiency on infrastructure, which includes servers and databases. Nowadays, the operating system, database and web application server can easily obtain as open source. However, as economies of scale issue; cloud technologies offer the chance of lower cost of infrastructure. Companies have noto maintain related environments for tasks such as building, configuring, testing, and deploying applications. Thus, this increases efficiency, decreases production time, and reduces risk by simplifying the process. Database and scalability global file services are among the cloud applications and products that offer strong consistency guarantees.

The introducedof cloud computing has bring a number of advantages not only for individuals, but also for businesses around the world. The first common benefit of using cloud computing is that it helps to reduce the cost in terms of time, money and storage. Cloud computing reduce costs in terms of money because it allow users to access their files from any personal computer especially the Software as a Service (SaaS) application, which enablesusers to use their current computer rather than buying a new one. Furthermore, cloud computing is easier to retain compare to the traditional computing since they use software that requires little involvement. Thus, ithelps to reduce the businesses' expenditure in terms of paying for hard drives, RAM, operating systems and hiring workers for IT department.

However, cloud computing helps to improve work from being time-consuming as we are able to access the file we need in a shorter time than finding the file manually. Furthermore, cloud computing can store more files than a personal computer can store since they stored it on a remote server, thus there is no more need to rent a physical space and buy and maintain servers and databases. Moreover, since most of the cloud service providersaccessedthe data at multiple times, therefore if one data centre could not access thesedata, it can still be accessible from other data centres. Reliability of cloud computing helps the company to prevent any IT emergencies after working hours, vendor management and procurement cycles. Anotherbenefit of using cloud computing would be in terms of accessibility. Cloud computing gives the same benefit like other types of computing where it allows users to access their files from anywhere as long as there is an internet and network connection of computer access. This benefit allows the users to have a greater sense of freedom as they can access the data from many devices such as mobile phones, tablets and even traditional computers. Furthermore, automatic update for the server will be an advantage for clients, theydo not need to hire people to update the server as it automatically update and once the server is updated, clientswill be able to use the updated services without downloading anything.

It is widely known that businesses minimized their costs toincreases their sales and profits. The ideas for them to get maximum profit is through business application software. However, businesses application softwareare too expensive. For example, businesses need a data centre with office space, power, cooling, bandwidth, network, servers and storages. Furthermore, they also need a complicated complex software stack and teams of experts to run deal with. The software needs development, testing, staging, and production. Such complex software usually creates problem where even technicians cannot fix it. Software always came up with updates of new version where users need to upgrade them but the consequence is that it may bring the whole system down.

As an example is Gmail as user as we do not need server-storage,and technicians team and upgrades it application. In cloud computing, businesses only have to log in, customize it and start using it. However, businessapplication software that are known as"Enterprise Cloud Computing" is running all kinds of apps in cloud computing (such as HR, Accounting, Etc). In cloud computing, security and new enhancement are up-to-date where cloud service provider will update it automatically. As for the payment, with cloud computing you do not have to buy anything,likesoftware and servers. It only depends on the predictable monthly subscription where businesses will only pay for what they used.

THE DRAWBACKS OF CLOUD COMPUTING: SECURITY

Despite the advantages that cloud computing brings, we have to bear in mind that every technology has its own drawbacks. The first common issue for cloud computing would be their security and privacy problem (Susanto et al, 2011). As we all know, cloud computing is about storing their files withthird party. For individuals, they might feel uneasy about sharing their files with another party especially the sensitive issues. Furthermore, as cloud computing allows the files to be accessed from any personal computer throughinternet connection, therefore viral infection and malware could occur (Susanto, Almunawar and Tuan, 2011).The second drawback would be in terms of control. Since cloud service providers offer to save client's files and documents, thus it is not reassuring to use since the data saved might be lost. As an exampleis experienced by T-Mobile's Sidekick smart phone services where in October 2009; the data of the clients have been lost and were not able to be retrieved. Therefore, it is quite risky to use cloud computing as there would be loss of control over the saved data.People are often worried about the security and reliability of cloud computing, surprisingly, compared to other network, cloud computing is more secured since most of the cloud service vendors utilisethe highest security certifications. However, It is imperative for cloud service provider and user to use an information security management system (ISMS) to effectively manage their information assets. ISMS is basically consist of sets of policies put place by an organization to define, construct, develop and maintain security of their computer based on hardware and software resources. These policies dictate the way in which computer resources coulduse.

Since information security has a very important role in supporting the activities of the organization through cloud network, we exteremely need a standard or benchmark which regulates governance over information security, these policies and standard as function as fundamental guidelines forcorporate secure electronic commerce on the global scale (Susanto et al, 2011b, 2012b). There are several standards for IT Governance which leads to information security awareness such as PRINCE2, OPM3, CMMI, P-CMM, PMMM, ISO27001, BS7799, PCIDSS, COSO, SOA, ITIL and COBIT. Unfortunately, some of these standards are not well adopted by the organizations, with a variety of reasons. The big five of ISMS standards, ISO27001, BS 7799, PCIDSS, ITIL and COBIT, widely used standards for information security is discussed and compared (figure 4) (Susanto, Almunawar, and Tuan 2012a)to determine their respective strengths, focus, main components and their adoption based on ISMS.

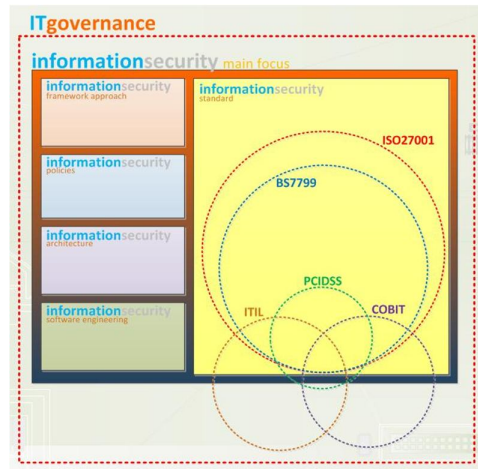

*Figure 2. Position of information security standard*

In conclusion (Susanto, Almunawar, and Tuan 2012a) each standard playsits own role and position in implementing ISMS, several standards such as ISO 27001 and BS 7799 focusing on information security management system as main domain and their focus on, while PCIDSS focus on information security relating to business transactions and smart card, then ITIL and COBIT focuses on information security and its relation with the project management and IT Governance (figure 2). However, refers to the usability of standards in global, indicated that ISO (27001) leading than four other standards especially on ISMS, therefore it is indicated that the standard is more easily implemented and well recognized by stakeholders (top management, staff, suppliers, customers/clients, regulators), the standard introduces a cyclic model known as the "Plan-Do-Check-Act" (PDCA) model, aims to establish, implement, monitor and improve the effectiveness of an organization's ISMS (Susanto, Almunawar & Tuan 2011b), thus compliance with information security standard, ISO 27001, is highly recommended with a variety of reason mentioned.

There are certain problems faced by the cloud computing users. In terms of connectivity, users could access their data and information only if internet connection is established. Unfortunatelly, cloud computing might caused major problems to its users as a lose tendency at certain time. An applied a higher level of security towards their data and confidential information is extremely needed (Susanto, Almunawar and Tuan, 2011). Therefore,implemented poor level of security, posibly that information stored on cloud computing exposed by unauthorized user and hackers (figure 3). Under cloud computing, once the data is transfer, it is no longer underuser's control,user haslostcontrol to the cloud sever.

It is believed that cloud computing is still considered as new. It is impossible to transfer a huge amount of data at its current performance status. Internet connection is essential for cloud computing, moreover, the internet connection will affect the quality of cloud computing.In the cloud service provider point of view, cloud-computing leads to higher cost especially, it would include a high amount of expenses to do research and make an improvement at the current cloud services that they offer. They would also have to invent software that will run in the cloud, rewire the equipment's and fix the unanticipated problems. In addition, client that use private cloud computing services would be paying a higher cost compared to the other cloud computing, public and hybrid cloud computing, since they used private computing infrastructure for them and not share with others to ensure their security.

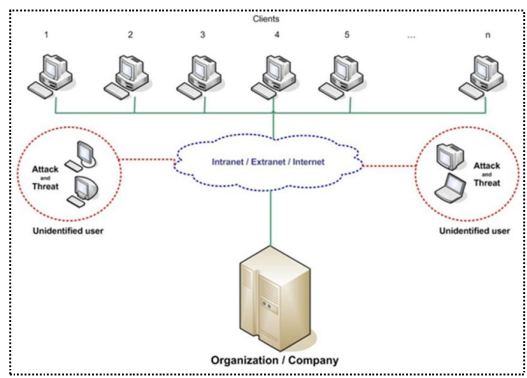

*Figure 3*.Activities of unidentified user as potential attack and threat to organization

## THE GROWTH OF CLOUD AND THE NEED OF TECHNOLOGY: TRENDY, DEMAND AND FUTURE

Cloud computing had created a new way of processing and organizing data which causedbusiness processes changed. Few years back before cloud computing, employee will need to access data through specific department where bunch of files are kept and organized. This kind of process usually takes a lot of time and is very costly because they will need someone to organize their data and process it. In this way, coordination among different section of an organization is better because information shared and passed through the cloud therefore all department managers can move at the same pace. Etro(2011) was able to improve coordination by using cloud computing which reduce their costs and enhances communication among its employers.As cloud computing is getting more demanding because of its contribution to the world and this provides more business opportunity for cloud computing service provider businesses. Lamba and Singh (2011) found out that Kalgidhar Trust created an education cloud based on its strong domain knowledge and the education. Cloud computing has brought e-commerce to a new generation where web hosting provided by third party and data storage is no longer a problem.  However, it becomes much cheaper to start an online shopping website,sincethere is noneed to build their own data centre, maintain the traffic of their website. Aljabre (2012) statedthat Amazon.com offers cloud storage service,alsoSaleforce's main product is a customer relationship management web service (Delgado, 2010).

Furthermore, cloud computing provide a great assistance in knowledge management because it is the solution of traditional learning and teaching in terms of passing knowledge and also sharing of knowledge. Furthermore, cloud computing mightreduce cost in education,since thereis noneed of any extra IT staff to maintain data centreand hardware maintenance. (Singh &Sodhi, 2012) said thatcloud will act as a core system to combine each institution to coordinate with each other therefore education cloud will contain the latest education and e-learning  software. There are many reasons why an individual should know and understand fully about cloud computing in this century where knowledge about Information Technology (IT) is vital to almost all industries. Thus, Cloud computing is not a stranger to IT but it gains popularity as more and more people depend on their smart devices, computer or tablets to store important documents instead printing it. Hence, as the growth of cloud computing is ascending from time to time the need to enhance technology are also essential in order to protect its users from any harm it may come across. Most people are familiar with bits of cloud computing which are Google docs and mail to send and receive emails, however a huge growth in cloud computing will occured,it has been stated by Forrester that cloud services will be delivered over the public internet in a small portion will generate $15.9 Billion revenue by 2020. Thus, this will involve companies licensing software that allow to easily sharing tech resources.

Microsoft predicted that cloud-computing marketwouldincreasenearly 14 million new jobs worldwide by 2015 while gaining revenue of $1.1 Trillion per year by 2015. It has signed a contract with the Indian government to run its cloud computing technology for the country's massive education network. Today's factor that contributes to the growth of cloud computing is due to the availability of low cost computing and bandwidth in the long run which therefore allows cloud hosting providers such as EarthLink Cloud to offer robust cloud services that help businesses save money. The idea of cloud service will give businesses a lower administration cost and prevent them from buying hardware that cost a lot, attracts all businesses to invest themselves with cloud computing, although they know the initial start-up of cloud computing is expensive but in the long term they will gain economies of scale.

As more and more business starts to recognise the benefits by just using cloud computing, they can easily save a lot of money, for example, "Wise Group" a non-profit group that employs 600 workers and have a turnover of£34m annually saves £300,000 a year using the cloud. It is proven that it is not just theory the cloud could lessen businesses costs. Minnesotastate has completed its migration to Microsoft Cloud after a year of testing as the federal government signed a deal with Microsoft in September 2010 to move 33,000 workers.

In the future, a strict and well-built security is exteremely needed.There is an increasing trend of using cloud to store private confidential documents. Hence, to avoid security breach by hackers it has stated that the security will move to "centralized trust" where by the authority learns to manage identities within the enterprises. In addition, having a centralized data will become a key strategic advantage to all of its users. The reason is in the near future cloud computing will create huge databases in the sky that aggregate valuable information that anyone can use through a publicly accessible.

Nowadays, as smartphones are getting more popular, the device will have to become more powerful. With the rise of mobile computing and the reliance on clouds to support mobile applications, mobile devices will have more capabilities in the cloud, i.e. Apple's iCloud. However, as convenient it may seem to have cloud in Smartphone there are also costs to acknowledge. The future of cloud computing seems bright with giant companies purchasing smaller companies that provide cloud services such as Dell acquired EqualLogic Inc. in 2007 for $1.4 billion as the foundation for its data-storage product, while HP bought Ospware, a data centre automation start-up for $1.6 billion. These multinational coorporation know how influential cloud computing couldbe in the future, as futurologist Dr James Bellini said. "If you go forward to the 2020s a successful enterprise will probably have no chief executive, no headquarters and no IT infrastructure,"

## V. CONCLUSION REMARKS

The main benefit of cloud computing is the ability to reduce costs where effectiveness and efficiency can be both achieved but initial costs will be needed because not all devices is optimized to run Cloud computing application. This could influence the whole society and economies of the world. Furthermore, the influential power of social media or web 2,0 are all given by cloud computing where reader has a chance to communicate with writer or with other reader and the messages is not just a bunch of words but can be visual images. However, the main problem of cloud computing is about security and privacyissues. Therefore, future cloud computing need to have a well-built security to ensure every user's data is secured, as in the future they will be more cloud user.

The demand for fastest access to information is changing and increasing, therefore, the availability of cloud computing has made it easier for organizations to share and store related data and information with their stakeholders such as the co-workers, creditors and customers. Moreover, some advantages is gained from these services has increased the confidence level of the organization to use cloud computing. However, the business organizations need to get a clear understanding on how cloud computing works and determine the potential risks they might face in the future.Therefore, since cloud computing involves in keeping and sharing information, organizations need to make sure that the cloud computing service provider is professional to store confidential information. Finally, with the advancement in technology nowadays, the convenient, cost saving and faster services are highly required by customers and cloud computing is one of the services that could meet all these requirements.

## AUTHORS

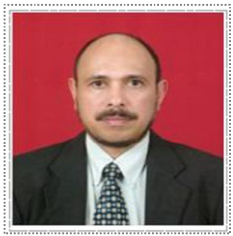

Mohammad Nabil Almunawar is a senior lecturer at Faculty of Business, Economics and Policy Studies, University of Brunei Darussalam. He received master Degree (MSc Computer Science) from the Department of Computer Science, University of Western Ontario, Canada in 1991 and PhD from the University of New South Wales (School of Computer Science and Engineering, UNSW) in 1997. Dr Nabil has published many papers in refereed journals as well as international conferences. He has many years teaching experiences in the area computer and information systems.

**Papers & Citations:**
*http://scholar.google.com/citations?user=AJy07UIAAAAJ&hl=en*

HeruSusanto is a researcher at The Indonesian Institute of Sciences, Information Security & IT Governance Research Group, also was working at Prince Muqrin Chair for Information Security Technologies, King Saud University. He received BSc in Computer Science from Bogor Agriculture University, in 1999 and MSc in Computer Science from King Saud University, and nowadays as a PhD Candidate in Information Security System from the University of Brunei.

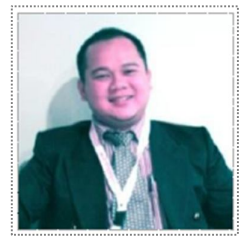

**Papers & Citations:** *http://scholar.google.com/citations?hl=en&user=rGgCYCoAAAAJ*

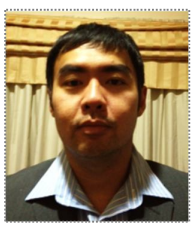

Chen Chin Kang is a lecturer at Faculty of Business, Economics and Policy Studies, University of Brunei Darussalam. He graduated with a Bachelors of Engineering (Hons), Royal Melbourne Institute of Technology in 1999 and continued with a Masters of Digital Communications, Monash University which was completed in 2001. He has worked in the private sector for a few before becoming a lecturer. In 2011, he completed another Masters of IT,Queensland University of Technology.